\documentstyle[epsfig,12pt]{article}
\begin{document} 
%
\begin{center}
{\large   }
\end{center}
\vspace{2 ex}
 
  
\begin{center}

\LARGE\bf
\rule{0mm}{7mm} Valence quark annihilation effect on charmed meson production in $\pi N$ collisions\\
\end{center}

\vspace{4ex}
\begin{center}
Tsutomu Tashiro$^1$, Hujio Noda$^2$, Kisei Kinoshita$^3$ and
Shin-ichi Nakariki$^1$ \\
\vspace{3ex}
$^1$Department Simulation Physics, Okayama University of Science, Ridai-cho, Okayama 700-0005, Japan \\
$^2$Department of Mathematical Science, Faculty of Science,
Ibaraki University, Bunkyou, Mito 310-0056, Japan  \\
$^3$Physics Department, Faculty of Education,
Kagoshima University, Korimoto, Kagoshima 890-0065, Japan\\
\end{center}
\vspace{2 ex}

\centerline{}

\vspace{1 ex}

\begin{abstract}

We discuss the hadroproduction of charmed mesons in the framework of 
the constituent cascade model taking into account the valence quark 
annihilation.  It is shown that the small valence quark annihilation process
dominates the leading particle production at large Feynman $x$ and
explains the recent experimental data on the asymmetry between $D^0$ and $\bar{D}^0$
at 350 GeV/c. \\

PACS \\
     13.75.Gx(Pion-baryon interactions)\\
     13.87.Fh(Fragmentation into hadrons)\\
     14.40.Lb(Charmed mesons)

\end{abstract}

\begin{center}
\end{center}
\vspace{1 ex}
\rule[.5ex]{16cm}{.02cm}
$^1$ email:  tashiro@sp.ous.ac.jp, nakariki@sp.ous.ac.jp, Fax: +81 86 256 8006 \\
$^2$ email:  noda@mito.ipc.ibaraki.ac.jp, Fax: +81 29 228 8407 \\
$^3$ email:  kisei@rikei.edu.kagoshima-u.ac.jp, Fax: +81 99 285 7735 \\

%
\setlength{\oddsidemargin}{0 cm}
\setlength{\evensidemargin}{0 cm}
\setlength{\topmargin}{0.5 cm}
\setlength{\textheight}{22 cm}
\setlength{\textwidth}{16 cm}
\setcounter{totalnumber}{20}
%
\pagestyle{plain}
\setcounter{page}{1}


\newpage
\section{Introduction}
\label{intro}
\indent

  Recently experiments at CERN \cite{wa92} measured the neutral $D$ mesons in $\pi^-$ nucleus
collisions and observed much smaller values of the leading/non-leading asymmetry 
than those of charged $D$ mesons i.e. less than 0.2 and even a negative value around $x=0.8$.
The experimental data on the asymmetry of charged $D$ mesons 
increases from zero to nearly one with Feynman variable $x$ in the $\pi^-$ fragmentation 
region \cite{wa92,wa82,e769,e791,e769_2}.  
The leading particle contains the same type of quark as one of the valence quarks in the 
incident hadron, while the non-leading one does not contain the projectile valence quarks. 
For example, the asymmetry of $D^0(c\bar{u})/\bar{D}^0(u\bar{c})$ in $\pi^-(d\bar{u}) $ 
interaction with nucleon is defined as
\begin{eqnarray}
A_{\pi^- N}(D^0,\bar{D}^0) = \frac{\sigma(D^0)-\sigma(\bar{D}^0)}{\sigma(D^0)+\sigma(\bar{D}^0)}.
\end{eqnarray}

In the perturbative QCD at leading order, the factorization theorem predicts that 
$ c $ and $ \bar{c} $ quarks are produced with the same distributions and 
then fragment independently.  In this case the asymmetry $A_{\pi^- N}(D^-,D^+)$ is equal to zero \cite{qs}.  Even 
in the case of next to leading order, the predicted asymmetry is much smaller 
than the data \cite{nde}. 
The asymmetry $A_{\pi^- N}(D^-,D^+)$ has been investigated and 
explained by means of many approaches: string fragmentation \cite{PYTHIA}, intrinsic charm 
contributions \cite{vb,anzivino,Armesto,piskounova,arakelyan_volko,arakelyan}, 
recombination process \cite{bednyakov,a_h_magnin_s,c_h_magnin}, recombination using
 valon concept \cite{hwa,das_hwa} and so on.
We have proposed the constituent quark-diquark cascade model and explained 
the leading/non-leading asymmetry of charged $D$ mesons $A_{\pi^- N}(D^-,D^+)$ 
successfully \cite{tnnik}.  The model, however,
gives rather large values of $A_{\pi^- N}(D^0,\bar{D^0})$ at $ 0.5 \stackrel{<}{\sim} x $ 
as expected from the leading particle effect but
deviating from the experimental data.  
Although several models have been applied more or less
satisfactorily to  $A_{\pi^- p}(D^0,\bar{D^0})$ at $ x \stackrel{<}{\sim} 0.6$
\cite{arakelyan,c_h_magnin}, the asymmetry problem about 
the charmed hadron productions is an open question. 

In the present paper
we investigate the leading/non-leading $D$-meson asymmetry in the framework of 
the constituent quark-diquark cascade model by taking into account the valence 
quark annihilation.

\section{Model description}
We consider an inclusive reaction $ A+B \rightarrow C+X$ in the centre of mass 
system of $ A $ and $ B$.  The light-like variables of $ A $ and $ B $ are defined 
as follows:
\begin{eqnarray}
x_{0\pm}^A=\frac{E^A \pm p_{cm}}{\sqrt{s_0}},~~~x_{0\pm}^B=\frac{E^B \mp p_{cm}}{\sqrt{s_0}} ,
\end{eqnarray}
where $ \sqrt{s_0} $ is the centre of mass energy of the incident hadrons $A$ and $B$.
We briefly review our model and then introduce the valence quark annihilation mechanism into
the model.

\begin{figure}
\resizebox{0.5 \textwidth}{!}{%
  \includegraphics{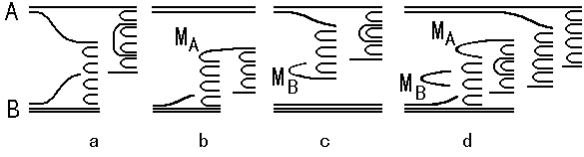}
}
\caption{The interaction mechanism in $AB$ collision: {\bf a} Non-diffractive dissociation type, 
{\bf b, c} Single-diffractive and {\bf d} double diffractive dissociation type mechanisms.}
\label{fgr:intrctn type}       
\end{figure}

 When the collision between $A$ and $B$ occurs, the 
incident hadrons break up into two constituents with a probability $ (1 - P_{gl})$; 
otherwise they emit wee gluons with $ P_{gl}$ followed by a quark-antiquark pair 
creation.   We assume four interaction types: a) non-diffractive dissociation, b) and c) 
single-diffractive dissociations of $ B$ and $A$, and d) double-diffractive dissociation types 
as shown in Fig. \ref{fgr:intrctn type}.  The probabilities of these types to occur 
are $ (1 - P_{gl})^2 , P_{gl} (1 - P_{gl}), P_{gl} (1 - P_{gl})$ and $ P_{gl}^2$, 
respectively.  Here we denote the quark-antiquark pair emitted from $ A $ ($ B $) 
via the wee gluons as $ M_A $ ( $ M_B $).  The probabilities of $ M_A $ ( $ M_B $) 
to be $u\bar{u}, d\bar{d}, s\bar{s} $ and $ c\bar{c} $  are denoted as 
$  P_{u\bar{u}},~ P_{d\bar{d}},~  P_{s\bar{s}} $ and $ P_{c\bar{c}}  $, respectively.  
%
%

The momentum fraction of $M_A$ is fixed by the distribution function
\begin{eqnarray}
  H_{M_A/A}(z)  = z^{\beta_{gl}-1}(1-z)^{\beta_{ld}-1}/B(\beta_{gl},\beta_{ld}) ,
\label{eqn:Hmaa}
\end{eqnarray}
and the uniform distribution $R$ in the interval from zero to one as,
\begin{eqnarray}
x_+^{M_A}=x_{0+}^Az,~~~x_-^{M_A}=x_{0-}^AR.
\label{eqn:Hmaa_2}
\end{eqnarray}
Then the incident particles $A$ and $B$ 
have the following momentum fractions:
\begin{eqnarray}
x_+^{A}=x_{0+}^A(1-z),~~x_-^{A}=m_A^2/(x_+^{A}~s_0),
\nonumber
\end{eqnarray}
\begin{eqnarray}
x_-^B=x_{0-}^B-(x_{-}^A-x_{0-}^A(1-R)),~~x_+^B=x_{0+}^B,
\label{eqn:Hmaa_2_leading}
\end{eqnarray}
where the mass shell condition is considered and transverse momenta are neglected. 
The momentum fraction of $M_B$ is treated similarly, exchanging the role of $A$ and $B$.
\\ 
 In the centre of mass system of incidents $A(M_A) $  and $ B(M_B)$, we define the 
light-like fractions of these hadrons and fix the light-like fractions of the 
projectile constituents.
The distribution functions of the constituents in the projectile $ A $ 
composed of $ a $ and $ a' $ are described as
\begin{eqnarray}
 H_{a/A}(z) = H_{a'/A}(1-z) = \frac{z^{\beta_a-1}(1-z)^{\beta_{a'}-1}}{B(\beta_a,\beta_{a'})}. 
\label{eqn:Ha/A}
\end{eqnarray}
Then the light-like fractions of $ a $ and $ a' $ are $ ~~x_+^a=x_{0+}^Az~,\\
 x_-^a=x_{0-}^AR,\ x_+^{a'}=x_{0+}^A-x_+^a$ and $x_-^{a'}=x_{0-}^A-x_-^a $, respectively.
The distribution functions of the constituents in $M_A, B$ and $M_B$ are similarly defined.

In our model hadrons are produced on the chain between a valence quark(anti-quark) 
from $A(M_A)$ and the valence diquark(quark) from $B(M_B)$ via the cascade processes

\noindent
\begin{eqnarray}
q & \rightarrow & M(q \bar{q}')+q', \nonumber \\
& & B(q[q'q''])+\overline{[q'q'']} ,  B(q\{q'q''\})+\overline{\{q'q''\}} , \nonumber \\  
\overline{[q'q'']} & \rightarrow & \overline{B}(\bar{q}\overline{[q'q'']})+q, \nonumber \\
& & M(q\bar{q}')+\overline{[qq'']} , M(q\bar{q}')+\overline{\{qq''\}}, \nonumber  \\ 
\overline{\{q'q''\}} & \rightarrow & \overline{B}(\bar{q}\overline{\{q'q''\}})+q, \nonumber \\
& & M(q\bar{q}')+\overline{[qq'']} ,M(q\bar{q}')+\overline{\{qq''\}},\nonumber\\
\end{eqnarray}
where $[\ ]$ and $\{\}$ denote the flavour antisymmetric and symmetric 
diquarks, respectively \cite{qdq}. 
 Meson production probabilities 
from $q, \overline{[q'q'']} $ and $ \overline{\{q'q''\}}$ are $ 1-\epsilon, \eta_{[~]} $ 
and $ \eta_{\{\}} $, respectively.

We redefine the light-like fractions of the incident constituents in the rest frame of the cascade chain.  The momentum sharing of the cascade process $ q + \bar{q}' \rightarrow M(q\bar{q}'') + q''+ 
\bar{q}' $ from a $q $ with $ x_\pm^q $ and $ \bar{q}' $ with $ x_\pm^{\bar{q}'} $ takes place as follows \cite{qdqA,km}:  First, using the emission function
\begin{eqnarray}
 F_{q''q}(z) = z^{\gamma\beta_q-1}(1-z)^{\beta_q+\beta_{q''}-1} /B(\gamma\beta_q,\beta_q+\beta_{q''}) , 
\label{eqn:Fqq} 
\end{eqnarray}
we fix the lightlike fractions of $ q'' $ and $ M $ as $ x^{q''}_{+} = x^q_{+} z $ and $ x^{M}_{+} = x^q_{+} - x^{q''}_{+} $, respectively and put $ x^{q''}_{-} = x^q_{-} $.  Second, the transverse 
momentum of $ M $ is determined from the probability function
\begin{eqnarray}
   G(\mbox{\boldmath$p$}_{T}^2)=\frac{\sqrt{m}}{C}\exp(-\frac{C}{\sqrt{m}}\mbox{\boldmath$p$}_{T}^2)
\label{eqn:pT2} 
\end{eqnarray}
in $ p_{T}^2 $ space. Then, from the onshell condition, $ x^{M}_{-} $ is fixed as  $ x^{M}_{-} = (m_M^2 + {{\mbox{\boldmath$p$}}_{T}}^2)/x^M_{+}s'$, where $ \sqrt{s'} $ is the subenergy of the cascade chain. The transverse momentum of $ q'' $ is $ {\mbox{\boldmath$p$}}^{q''}_T = {\mbox{\boldmath$p$}}^q_T - {\mbox{\boldmath$p$}}_{T} $.  The lightlike fraction of $ \bar{q}' $ is decreased to $ \tilde{x}^{\bar{q}'}_{-} = x^{\bar{q}'}_{-} - x^{M}_{-}$.  If the energy of $\bar{q}'$ is enough to create another hadron, the cascade such as $ q'' + \bar{q}' \rightarrow q'' + \bar{q}''' + M(q'''\bar{q}') $ takes place in the opposite side.
Finally recombined hadrons are put on-shell by a two body decay process 
as explained in \cite{tnnik}.

The dynamical parameters $ \beta$'s in (\ref{eqn:Ha/A}) and (\ref{eqn:Fqq}), which 
determine the momentum sharings of the constituents, are 
related to the intercepts of the Regge intercepts as $
\beta_u=\beta_d=1-\alpha_{\rho-\omega}(0),~~ \beta_s=1-\alpha_\phi(0),
\beta_c=1-\alpha_{J/\psi}(0)$ \cite{mnkt,cthkp}. 
From previous analyses \cite{qdq,qdqA}, we determine the values for diquarks as
$\beta_{[ij]}=\gamma_{[~]}(\beta_i+\beta_j),~~\beta_{\{ij\}}=\gamma_{\{\}}(\beta_i+\beta_j)$.
We consider lower lying hadrons: pseudoscalar($PS$), vector ($V$), tensor($T$) mesons,
 octet($O$) and decuplet($D$) baryons composed of $ u,d, $ and $ s $ flavours and 
the correspondings with charm flavour.  We assume the production 
probabilities for them to be $ P_{PS},  P_V,  P_T $ ($=1-P_{PS}-P_V),
 P_O $ and $ P_D$ ($=1-P_O$), respectively.  Octet 
and decuplet baryons are described as 
\begin{eqnarray}
|8 > = \cos\theta|q[q'q'']> + \sin\theta|q\{q'q''\}>,
\label{eqn:octB} 
\end{eqnarray}
\begin{eqnarray}
|10 > = |q\{q'q''\}>.
\label{eqn:dcpB} 
\end{eqnarray}
Directly produced resonances decay into stable particles. Details
of our model are explained in \cite{tnnik,qdq,qdqA}.

We modify our model to include valence quark annihilation 
in the non-diffractive dissociation type interaction.
Let us consider the case of $\pi^-p$ collision.
We take account of the annihilation of the valence $\bar{u}$ from $\pi^-$ and valence $u$ 
from proton target in a slightly different way from the one pointed out in 
\cite{bednyakov,c_h_magnin}.  
There are annihilation processes such as shown in Fig.2.  The annihilation process in Fig.2a is 
considered soft process and its contribution is related to the magnitude of $\sigma_{inel}^{\pi^-} 
- \sigma_{inel}^{\pi^+}$ by unitarity.  This process, however, may be negligible for the charm-pair production.
Here we only take into account the annihilation process in Fig.2b and
 assume that the process
\begin{eqnarray}
\bar{u}u \rightarrow \bar{q}q
\label{eqn:anni} 
\end{eqnarray}
occurs with 
the probability $P_{anni}(1-P_{g\ell})^2$ and the non-diffractive type occurs with 
the probability $(1-P_{anni})(1-P_{g\ell})^2 $. Branching ratios of $\bar{u}u \rightarrow 
\bar{u}u, \bar{d}d, \bar{s}s,$ and $\bar{c}c$ are chosen to be equal to each other 
for the channels allowed energetically. 
%
\begin{figure}
\resizebox{0.47\textwidth}{!}{%
  \includegraphics{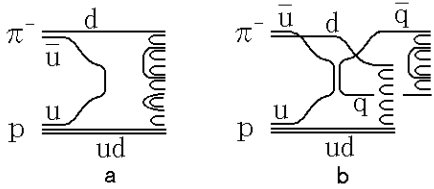}
}
\caption{The valence quark annihilation processes {\bf a} soft
annihilation and {\bf b} semi-hard annihilation processes in $\pi^- p$ collision.}
\label{fgr:vqanni}       
\end{figure}

 This process is considered as a semi-hard scattering 
process and produced $q$ and $\bar{q}$ are supposed to be non-free.
It seems natural to assume that $\bar{q}$ has tendency to be
 produced in the forward direction of the $\bar{u}$ 
in the centre of mass system of $\bar{u}$ and $u$, 
due to the confinement force between 
the valence $d$ quark and the produced $\bar{q}$ quark.
 Here we choose the distributions for $z=\cos\phi$ as 
\begin{eqnarray}
D(z)=\frac{3}{8}(1+z)^2 
\label{eqn:annidrctn} 
\end{eqnarray}
in the region $-1 < z < 1$, 
where
$\phi$ is the angle between the directions of $\bar{u}$ and $\bar{q}$
in the centre of mass system of $\bar{u}$ and $u$.
 After this 
annihilation mechanism, the non-diffractive type production of hadrons occurs as shown in Fig. 2b.

\section{Comparison with the data}
\label{sec:3}
In this section, we give the results of our model for the inclusive 
hadron productions in $\pi^- N$ collisions.  The value of the 
dynamical parameter $\beta_c$ is changed from the value $\beta_c=8.0$ used in \cite{tnnik} 
to $\beta_c=4.0$.  This is in accord with the argument that the slopes of Regge trajectories
of charmed mesons are smaller than those of light mesons as discussed in \cite{Burakovsky_g}. 
The parameter of $p_T^2$ distribution $ C=1.2$ and the probability 
$P_{c \bar{c}}=0.00016 ( P_{s\bar{s}}=0.09984)$ in \cite{tnnik} are changed into
$C=1.0$ and $P_{c \bar{c}}=0.00025 ( P_{s\bar{s}}=0.09975)$.  
Furthermore the meson production 
probabilities are assumed as $P_{PS}=0.4, P_{V}=0.4 $ and $P_T=0.2$ for light mesons 
and $P_{PS}=1/9, P_V=1/3$ and $P_T=5/9$ for charmed mesons.
 We put the parameter $P_{anni}$ 
so as to reproduce the asymmetry $A_{\pi^-N}(D^0,\bar{D^0})$.  We choose 
the value $P_{anni}=0.0005$. This value scarcely changes the features of the spectra of 
light hadrons with $u, d$ and $ s$.   For other parameters, we use the same values 
used in the previous analysis \cite{tnnik}.

In Fig.3 we show the results of $D^{*}$ productions and compare the two cases:
case(1) $3(1+z)^2/8$ distribution for $\cos \alpha $ and case(2) without annihilation ($P_{anni}=0$:the same as 
in \cite{tnnik} except for the values of $\beta_c=4.0, C=1.0, P_{c\bar{c}} = 0.00025$, 
and $ P_{PS}=1/9, P_V=1/3$ and $ P_T=5/9$ for charmed meson productions). The annihilation effect 
of Fig.2b is seen in $\bar{D}^{*0}(u\bar{c})$ and $D^{*-}(d\bar{c})$.
Our model gives a satisfactory description of $x$ dependence for production of
 $ D^{*+}$ or $D^{*-}$ in
$\pi^-N$ collision.  The result for production of $D^{*0}$ or $\bar{D}^{*0}$ at $x \approx 0.1$ is small
as compared with experimental data \cite{aguilar}.

The annihilation effect on meson 
productions in $\pi^- p$ is twice as much as in $\pi^- n$ collision.
The annihilation mechanism
have a considerable effect on the non-leading particle $\bar{D}^{*0}(u\bar{c})$.
However, the shape of the leading particle $D^{*-}(d\bar{c})$ is affected little 
by the annihilation mechanism.  Therefore our model gives larger differences
between $A_{\pi^- p}(D^{*0},\bar{D}^{*0})$ and $A_{\pi^- n}$ ($D^{*0},\bar{D}^{*0})$
than those between $A_{\pi^- p}(D^{*-},D^{*+})$ and $A_{\pi^- n}(D^{*-},D^{*+})$.
  Fig.4a shows the results in case(1) for $A_{\pi^- p}(D^{*0},\bar{D}^{*0})$,
$~A_{\pi^- n}(D^{*0},\bar{D}^{*0})$ and $A_{\pi^- N}(D^{*0},\bar{D}^{*0})$. 
 Hereafter we show 
the average results of $\pi^-$ beam on proton and neutron targets.
  Fig.4b shows the results of $A_{\pi^- N}(D^{*0},\bar{D}^{*0})$ and $A_{\pi^- N}(D^{*-},
D^{*+})$ in case(1) and case(2). 

Smaller values of $A_{\pi^- N}(D^{*-},D^{*+})$ as compared with 
$A_{\pi^- N}(D^{*0},\bar{D}^{*0})$ at $x\approx 0.5$ are due to 
the difference between the hadron productions on quark-diquark and antiquark-quark chains 
in non-diffractive dissociation type mechanism (Fig.1a) \cite{tnnik}. 
The leading particle  $D^{*-}(d\bar{c})$ is produced on the chain between the valence 
$d $ quark in the beam and the valence diquark in the target.  
The proton target has 
a tendency to break into an energetic valence diquark and a wee valence quark.
  Then the $ D^{*-}$ meson produced in 
the first cascade step is energetic.  However, in the case in which a baryon 
is produced from the valence diquark in the first cascade step, the momentum 
of the valence $d$ quark is decreased. Furthermore the total momentum is shifted to 
the diquark side.  Then the leading spectrum of $D^{*-}$ on the quark-diquark chain tends 
to have a small momentum and the asymmetry is reduced in 
$0.3 \stackrel{<}{\sim} x \stackrel{<}{\sim} 0.7$. 

On the other hand, the leading 
particle $D^{*0}(c\bar{u})$ is produced on the chain between the valence $\bar{u} $ quark 
in the $ \pi^- $ beam and the valence quark in the target. 
The momentum reduction of the $\bar{u}$ quark is small even in a case a meson is produced in the 
first cascade step from the valence quark in the target fragmentation. 
Therefore the 
asymmetry $ A_{\pi^- N}(D^{*0},\bar{D}^{*0}) $ is not so suppressed as compared with
 $ A_{\pi^- N}(D^{*-},D^{*+})$ in 
$ 0.3 \stackrel{<}{\sim} x \stackrel{<}{\sim} 0.7$. 
In case(1), the annihilation effect compensates the quark-diquark effect and the asymmetry
$A_{\pi^- N}(D^{*0},\bar{D}^{*0})$ becomes negative at large $x$. 

In Fig. 5, we show the result of $x$ dependence of $D$ mesons in 
$\pi^- N$ collisions for case(1) and case(2). The annihilation effect is also
seen in $\bar{D}^0$ and a little in $D^-$ spectra as in $D^*$ spectra.
In Fig. 6, we show the result of $p_T^2$ dependence of $D$ mesons in 
$\pi^- N$ collisions for case(1) and case(2).  
Our calculation shows smaller values of 
$ D^0$ than those of $\bar{D}^0$ at large $p_T^2$ deviating from the data.
 The newly introduced annihilation process 
explains the large $p_T^2$ charmed meson productions in part.  Our model is in good agreement with 
the data except for $D^0,~\bar{D}^0$ at large $p_T^2$. 
In Fig. 7 the results of $x$ dependence of $A_{\pi^- N}(D^-,D^+)$ and
$ A_{\pi^- N}(D^0,\bar{D}^0)$ are compared with the experimental data 
at 350 GeV/c \cite{wa92}. 
The agreement of $x$ dependence with the experimental data is satisfactory. 
The negative value of $ A_{\pi^- N}(D^0,\bar{D}^0)$ at large $x$ is well
explained by introducing a small amount of the valence quark annihilation process 
to the model.  Fig.8  presents the 
results of $p_T^2$ dependence of $A_{\pi^- N}(D^-,D^+)$ and $ A_{\pi^- N}
(D^0,$$\bar{D}^0)$.
The $p_T^2$ dependence of $ A_{\pi^- N}(D^0,\bar{D}^0)$ disagrees with the data in sign
as seen in Fig.6b.   The discrepancy increases with the annihilation effect.
The results of the $x$ and $p_T^2$ dependences of $A_{\pi^- N}(D^-,D^+)$ and
$ A_{\pi^- N}(D^0,\bar{D}^0)$ at 500 GeV/c are compared with the experimental 
data \cite{e791} in Fig.9. The agreement with the data is fairly good except for
the negative value of the data at $x \approx -0.15$. 
There is a valence $d$ quark
in the incident nucleon and our model predicts a positive value of $A_{\pi^- N}(D^-,D^+)$
for $ x < 0$ as seen in Fig.9.  However, in the case of proton target with $\cos \theta =1$ 
in (\ref{eqn:octB}), there is no valence $d$ quark and the model gives a small value of $A_{\pi^- N}(D^-,D^+)$
 in the target fragmentation region.

The results of the $x$ 
and $p_T^2$ dependence for production of $ D_S^-(s\bar{c})$ or $ D_S^{+}(c\bar{s}) $
are compared with the experimental data \cite{wa92} in Fig.10. The features of the spectra are in good agreement
with the data except for the small discrepancy in normalization.  Fig.11 shows the calculated results 
of asymmetries $A_{\pi^- N}(D_S^-,D_S^+)$ with respect to $x$ and $p_T^2$. The values are small
because both $D_S^\pm$ mesons are non-leading particles.  But in case(1) $A_{\pi^- N}(D_S^-,D_S^+)$
increases with $x$ at large $x$ due to the valence quark annihilation.

\section{Discussions}
\label{sec:4}
The large leading/non-leading asymmetry $A_{\pi^- N}(D^-,D^+)$ is naturally explained by 
our constituent quark cascade model as in the case of light hadron productions. 
In \cite{tnnik}, we noticed that 
the cascade chain properties are different between the antiquark-quark
and the quark-diquark chains.   The leading particle $D^-(d\bar{c})$ produced in the quark-diquark 
chain in $\pi^- p$ collision is less energetic when a baryon is produced in the target fragmentation
in the first cascade process as compared with $D^+(c\bar{d})$ produced in the 
quark-antiquark chain in $\pi^+ p$ collision. 
 We have different leading particle effects between  
$ \pi^+ $ and $ \pi^- $ beams i.e. $A_{\pi^+ N}(D^+,D^-) > A_{\pi^- N}(D^-,D^+)$ 
 around $x \sim 0.6 $ \cite{tnnik}.  Although the difference between the asymmetries 
$A_{\pi^- N}(D^0,\bar{D}^0)$ and $A_{\pi^- N}(D^-,D^+)$ disappears
due to the decay effect and the valence quark annihilation effect,this quark-diquark chain effect 
is seen in the difference between the asymmetries $A_{\pi^-}(D^{*0},\bar{D}^{*0})$ and 
$A_{\pi^- N}(D^{*-},D^{*+})$ \\ around $ x \sim 0.5 $ as shown in Fig.4b.

 The small valence quark annihilation process (\ref{eqn:anni}) explains the 
negative value of $A_{\pi^- N}(D^0,\bar{D}^0)$ at large $x$ observed in the 
recent experiment \cite{wa92}.  
This process is considered as 
semi-hard interaction.  All channels occur with equal probability and the
produced constituents tend to have the forward direction of the incident valence constituents.
This is a part of the hadron production process at large $p_T^2$.  Our model can explain the
spectra up to the region $ p_T^2  \stackrel{<}{\sim} 10$ (GeV/c)$^2$.
As noticed in the previous section, the results of calculated behaviours of 
the $p_T^2$ dependence of $A_{\pi^- N}(D^0,\bar{D}^0)$ disagree with the experimental data.
This implies that there may be considerable contribution from soft interactions which 
maintain the leading particle effect even at
large $p_T^2$ region.  The negative values of experimental data on $A_{\pi^- N}(D^0,\bar{D}^0)$ 
at $ p_T^2 \stackrel{<}{\sim} 3$ (GeV/c)$^2$ suggests that the $\bar{q}$ quark in the 
annihilation process (\ref{eqn:anni})
is produced in the very small cone around the direction of incident $\bar{u}$ as compared with 
(\ref{eqn:annidrctn}). It is interesting
to investigate the leading/nonleading asymmetry in $K N$ collisions, since in $K^+ N$
collision, there is no annihilation process.
These points will be discussed elsewhere.

%
%
%


\clearpage
\begin{figure}
\caption{Comparison of the model with the experimental data on the $x$-dependences for productions 
of {\bf a} $D^{*+}$ or $D^{*-}$ and {\bf b} $D^{*0}$ or $\bar{D}^{*0}$ mesons at $P_L=360$ GeV/c \cite{aguilar}. 
 The theoretical lines were calculated for $P_L=350$ GeV/c, full line for case(1) with
 the valence quark annihilation and dotted for case(2) without the valence quark 
annihilation.} 
\label{fig:3}       
\end{figure}
\begin{figure}
\caption{{\bf a} The $x$-dependences of the asymmetries $A_{\pi^- p}(D^{*0},\bar{D}^{*0}), 
A_{\pi^- n}(D^{*0},\bar{D}^{*0})$ and $A_{\pi^- N}(D^{*0},\bar{D}^{*0})$.  {\bf b}
The comparison of $x$-dependence of the asymmetries $A_{\pi^- N}(D^{*0},\bar{D}^{*0})$ and
 $A_{\pi^- N}(D^{*-},\bar{D}^{*+})$ between in the case(1) and case(2).}
\label{fig:4}       
\end{figure}
\begin{figure}
\caption{The $x$-dependences of {\bf a} $D^\pm$ and {\bf b} $D^0$ and $\bar{D}^0$ mesons at
$P_L=350$ GeV/c.  The full lines were calculated for
case(1) with the valence quark annihilation and dotted for case(2) without
 the valence quark annihilation.  The experimental data are taken from \cite{wa92}.}
\label{fig:5}       
\end{figure}
\begin{figure}
\caption{Same as in Fig.5 for $p_T^2$-dependence in range $0 \leq x \leq 1$.}
\label{fig:6}       
\end{figure}
\begin{figure}
\caption{The $x$-dependences of asymmetries {\bf a} $A_{\pi^- N}(D^-,D^+)$ and {\bf b}
 $A_{\pi^- N}(D^0,\bar{D}^0)$ at $p_L=350$ GeV/c. The full lines show the results in case(1) and dotted lines
in case(2). The experimental data are taken from \cite{wa92}.}
\label{fig:7}       
\end{figure}
\begin{figure}
\caption{Same as in Fig.7 for $p_T^2$-dependences in range $0 \leq x \leq 1$.}
\label{fig:8}       
\end{figure}
\begin{figure}
\caption{The asymmetry $A_{\pi^- N}(D^-,D^+)$ at $p_L=500$ GeV/c {\bf a} as a function of $x$, {\bf b} as 
a function of $p_T^2$ for range of $ -0.2 \leq x \leq 0.8$ and  $ 0.4 \leq x \leq 0.8$.  
 The experimental data are taken from \cite{e791}.}
\label{fig:9}       
\end{figure}
\begin{figure}
\caption{Differential cross sections with respect to {\bf a} $x$ and {\bf b} $p_T^2$ for production of
 $D_S^+$ or $ D_S^-$ at $P_L=350$ GeV/c.  The full line denotes the results for case(1) 
and dotted for case(2).  The experimental data are taken from \cite{wa92}.}
\label{fig:10}       
\end{figure}
\begin{figure}
\caption{Asymmetries $A_{\pi^- N}(D_S^-,D_S^+)$ with respect to {\bf a} $x$ and {\bf b} $p_T^2$ for
 at $P_L=350$ GeV/c.  The full line denotes the results for case(1) 
and dotted for case(2).}
\label{fig:11}       
\end{figure}

\end{document}